Article

# SRAM-Based PUF Reliability Prediction Using Cell-Imbalance Characterization in the State Space Diagram

Gabriel Torrens [1,2,*], Abdel Alheyasat [1], Bartomeu Alorda [1,3] and Sebastià A. Bota [1,2]

1   Electronics Systems Group, Industrial and Construction Engineering Department, University of the Balearic Islands, 07120 Palma, Spain; heyasat.abdel@uib.es (A.A.); tomeu.alorda@uib.es (B.A.); sebastia.bota@uib.es (S.A.B.)
2   Biosensors, Medical Instrumentation and Data Analysis Group, Health Research Institute of the Balearic Islands (IdISBa), 07120 Palma, Spain
3   eHealth and Multidisciplinary Telemedicine Research Group, 07120 Palma, Spain
*   Correspondence: gabriel.torrens@uib.es

**Abstract:** This work proposes a methodology to estimate the statistical distribution of the probability that a 6T bit-cell starts up to a given logic value in SRAM memories for PUF applications. First, the distribution is obtained experimentally in a 65-nm CMOS device. As this distribution cannot be reproduced by electrical simulation, we explore the use of an alternative parameter defined as the distance between the origin and the separatrix in the bit-cell state space to quantify the mismatch of the cell. The resulting distribution of this parameter obtained from Monte Carlo simulations is then related to the start-up probability distribution using a two-component logistic function. The reported results show that the proposed imbalance factor is a good predictor for PUF-related reliability estimation with the advantage that can be applied at the early design stages.

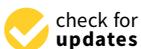

## 1. Introduction

Security is one of the key factors in current electronic devices, particularly in the Internet of Things (IoT) sector [1]. Some of the most important security requirements are trusted device identification methods and secured communication links between nodes.

Physical unclonable functions (PUFs) have turned out to be promising techniques for secure key generation and reliable identification. A PUF is capable of providing a response that can be used as a unique identifier. This response is said to be physically unclonable because it is obtained from a physical device feature that is virtually impossible to replicate [2]. PUFs are gradually being used as an alternative to the traditional non-volatile memories approach to store identifier keys. They show better attack resilience, flexibility, and, depending on the PUF type, also less cost [3]. These factors are particularly crucial in the IoT field, where security requirements can be high. However, many traditional security approaches are not suitable because IoT devices often have tight constraints in terms of cost or power consumption [4].

A good PUF has three attributes that should have a high score: uniqueness, unpredictability and reliability [2]. Uniqueness measures the differences between the responses from different PUF instances, unpredictability assesses if PUF response can be guessed, and reliability evaluates if a single PUF provides always the same response [5].

Many circuits have been proposed as PUFs, like ring oscillators [6], delay-based arbiter PUFs [7], flip-flops PUFs [8], and others [9–11]. The main drawback of these strategies is that they often require additional hardware, which in turn has an impact on cost and power consumption. Nevertheless, using Static Random Access Memories (SRAMs) as PUF has the advantage that most digital electronic devices include SRAMs, and thus, they can be used for both storing data and for PUF generation. For this reason, using SRAM as PUF is an option that can offer high performance at a low cost [12].

PUF-based SRAMs rely on the effect of random process variations in the transistor parameters that form the basic SRAM cell [13]. Differences in transistors parameters are originated during the manufacturing process and are random from cell to cell, and more importantly, from chip to chip. Therefore, these differences are unique and unpredictable, and thus, they can be used to uniquely and securely identify a chip. With data obtained from various cells of the same chip, a unique digital signature can be extracted. This signature will not only be exclusive for the chip, but also it will be virtually impossible to be predicted or replicated on another chip because it is based on random differences originated in the manufacturing process.

In SRAM PUFs, cell data is usually extracted from the start-up value of each cell, i.e., after a period where the supply voltage has been turned off [14]. Then, the cell is powered-up and its content is read-out. A cell is suitable to be used as PUF if its start-up value is the same every time it is started. Consequently, it is important to know the number of cells in a chip that will always start up to the same value and, thus, how many of them will be the best to be used as PUF.

The objective of this paper is to develop a methodology to predict the statistical distribution of cells in a given SRAM as a function of their probability to start up to the same value. Those with a high probability will be the most appropriate to implement PUFs. The prediction is made from an electrically simulated parameter that measures cell asymmetry. However, these results do not directly provide the desired probability distribution. To achieve that, simulated data are converted into start-up probabilities by using a transfer function calibrated with experimental results.

The paper is structured as follows. Section 2 introduces the working principles of SRAM PUFs. Section 3 describes the used SRAM. Section 4 presents the start-up experimental results. Section 5 introduces the proposed imbalance factor. Section 6 describes the methodology to relate the imbalance factor with the experimental start-up behavior. And, finally, Section 7 concludes the work.

## 2. SRAM PUFs

SRAMs are one of the most common blocks in digital integrated circuits. In addition, they are usually designed with transistor sizes close to the minimum allowed by technology [15]. This fact, together with the aggressive scaling of CMOS technologies, make SRAMs particularly sensitive to parameter variations and external noise. This, in turn, can threaten SRAM stability and data integrity particularly in nanometer technologies [16,17]. Nevertheless, parameter variations are the basis for SRAM PUF generation, as it was introduced in the previous section, and as it will be further discussed in this section.

The most common SRAM cell types are formed by either six or eight transistors [18–21]. In this work, we focus on the six-transistor cell, the so-called 6T cell. The schematic of the 6T cell is depicted in Figure 1a and it is composed of two cross coupled inverters that form a latch (transistors $N_1$, $N_2$, $P_1$, $P_2$), and two pass transistors ($N_{X1}$, $N_{X2}$) which allow reading and writing data. The cell has two nodes (Q and QB in Figure 1a), and once the cell is stable, one of them is at $V_{DD}$ ('1' logic state), whereas the other is at 0 V ('0' logic state), so the cell has two stable states: ($V_{DD}$, 0) and (0, $V_{DD}$). Which node is '0' and which one is '1' determines the bit that the SRAM cell is storing. Furthermore, pass transistors are used to write new data to the cell and to read its content. These two tasks are performed through the bit-lines (BL and BLB).

When the cell is off ($V_{BIAS}$ = 0 V), both nodes Q and QB are at 0 V. When the cell is powered on, $V_{BIAS}$ is raised from 0 to nominal supply voltage ($V_{DD}$). After a transient time, one of the two nodes will end up at 0 V, while the other will end up at $V_{DD}$. This means that the cell will start up either at logic '0' or at logic '1'.

If the two cross-coupled inverters were identical and perfectly balanced, a single cell would start up at logic '0' or at logic '1' randomly. In this hypothetical situation, the only factors which could tip the scale to one of the two states will be non-deterministic influences like external or internal noise, small temperature imbalances or other factors that can be considered random in time. However, this scenario is not realistic, since there are always differences in the transistor parameters that result in an imbalance in the cross-coupled inverters of each cell. The relative differences between the inverters of each bit-cell make each individual cell more likely to start up either at logic '0' or logic '1' [22].

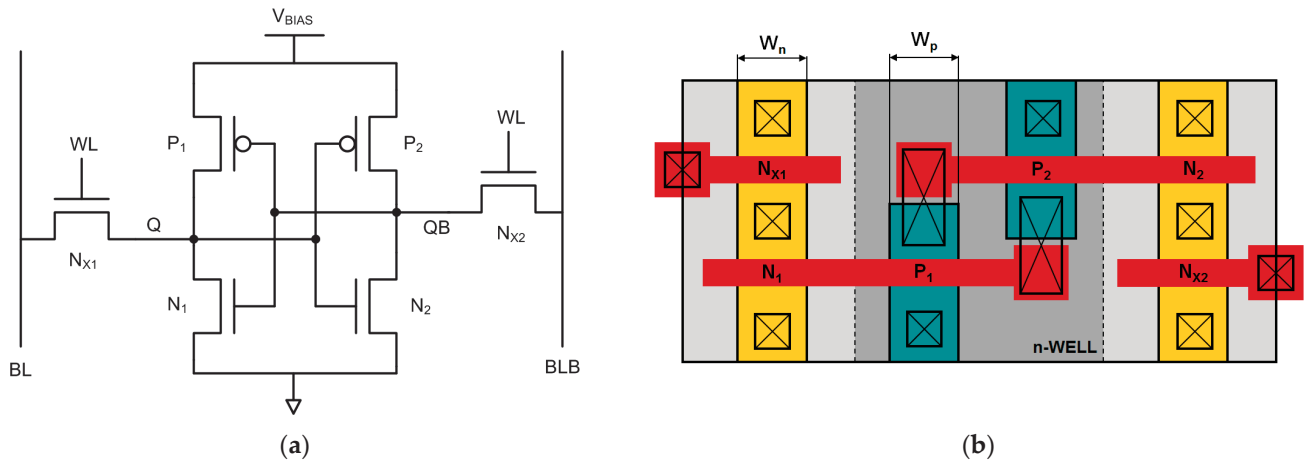

**Figure 1.** (**a**) Schematic of the six-transistor SRAM cell (6T cell) (**b**) Layout of the 6T cell without metal lines.

Highly mismatched cells will be more prone to start up to a given value each time they are powered up. These cells are highly biased to one of the two states, and they will always start to the same state, or at least be more prone to start to one given state because they will be less affected by external factors. By contrast, cells with near perfectly matched inverters will behave more randomly because, since they do not have an inherent bias towards any estate, they will be more affected by random factors like noise or temperature.

Highly mismatched bit-cells are those that are useful for PUFs because their start-up value will be more consistent between start-ups. These cells are usually called "reliable cells" from a PUF point of view. Conversely, well balanced cells are not suitable for PUFs because its response will be more unpredictable, and thus unreliable to be used as PUF. Nevertheless, unreliable cells can be used for random number generation [1,23,24].

SRAM cells designed to be used as memory storage are usually manufactured aiming to be as well balanced as possible to maximize their stability and dependability when storing data, when being read or when being written. This is in conflict with SRAMs working also as PUFs, since, as discussed above, the best SRAMs for PUFs are those which are more unbalanced. For this reason, if general-purpose SRAMs are to be used as PUF, it is necessary to implement some procedure to generate a reliable PUF response from a group of cells. There are mainly two methods to achieve this goal.

The first method is based on using error correction codes, repetition codes or temporal majority voting techniques. In the last case, the final output of a cell is obtained after challenging it several times and determining the most common challenge response [25]. The main drawback of this approach is that it is time consuming. In addition, the overall reliability of the result depends on how reliable the involved cells are. Under a certain individual reliability threshold, the result can be poor. Also, if error correction codes are used, usually implementing helper data algorithms (HDAs), the chances of a secure key being tampered are increased [12]. Furthermore, environmental conditions like temperature, aging or radiation can affect the result because they can modify cell reliability, either by worsening or even improving it [26–28].

The second method intends to obtain a more consistent response minimizing the effect of environmental conditions by using as PUF-cells only those cells that show a more constant behavior. These cells will be less susceptible to effects like aging, noise or temperature, and thus provide a more reliable overall result. To implement this method, it is necessary to identify the more reliable cells using an appropriate selection method. Once identified, only the most reliable cells are included in the PUF, while the others are masked out. The location of the suitable cells can be saved and reused [14,29]. However, this subset of cells may vary in time due to aging, and they may require revision.

This work focuses on this second method, which is compatible with memories that have not been specifically designed for PUF applications. Its aim is to develop a methodology to estimate the statistical distribution of cells that are suitable for being used as PUF as a function of its start-up probability. The proposed approach uses an imbalance factor obtained from Monte Carlo analysis to quantify the cell mismatch and relate the resulting distribution to start-up experimental data.

## 3. SRAM Characteristics

All the experimental data are obtained from SRAMs designed and implemented on a 65-nm commercial CMOS technology using standard Vt transistors with a nominal supply voltage of $V_{DD}$ = 1.2 V. The memory block contains a total of 16,384 cells, distributed in 256 rows and 64 columns. The used SRAM cell is the standard 6T-cell (Figure 1a). Its layout is a full-custom design, following the so-called structured or wide-layout design [30]. The layout structure can be seen in Figure 1b and the sizes of the transistors are $W_n$ = $W_p$ = 0.15 µm. This layout type provides better parameter variation control by applying design for manufacturing techniques, which include avoiding bends in the diffusion regions by having all nMOS and pMOS transistors of the same width, and setting all poly lines aligned and placed in the same direction [31]. The cell power supply can be controlled so that they can be shut down and set to nominal voltage. This makes it possible to power down the memory, power it up again, and read cell values. In Figure 2, a diagram of the structure of SRAM memory used in this paper is shown.

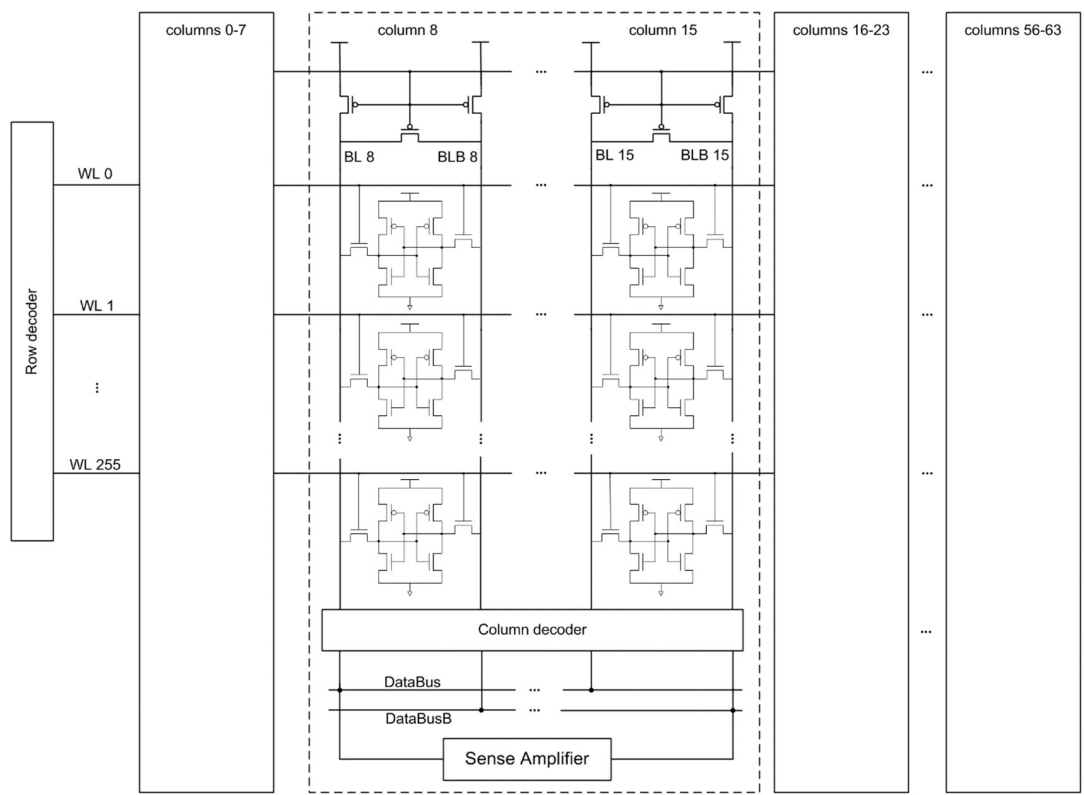

**Figure 2.** Diagram of the SRAM memory. It has a total of 16,384 cells, structured in 256 rows and 64 columns.

Cells are written and read through pass-transistors, which are controlled by a word-line signal (WL) common to all cells in the same row (Figure 2). When pass transistors are activated, cells are connected to the bit-lines (BL and BLB), which are common to all cells in the same column (Figure 2). Each set of 8 columns is connected to a sense-amplifier to read cell contents, so we have a total of 8 sense amplifiers for the 64 columns. The memory has a row decoder to allow selecting the actual row to be accessed. It also has a column decoder to select the column that needs to be connected to the sense amplifier, and thus read out. All control signals are generated from a control unit, which can be externally controlled by a logic analyzer. The memory uses words of eight bits to write or read data.

Simulations were performed with the Spectre simulator using Monte Carlo analysis to take into account parameter variations. Moreover, a set of 16,384 simulations were performed to have the same amount of data as from experimental results. Start-up process simulations require pass transistors to always be cut-off (for instance, simulations of the parameter proposed in Section 5). In this case, the cell is isolated from the rest of the circuit and can be simulated alone without affecting the results. We corroborated this hypothesis and performed all cell start-up simulations with one single cell. This is an important advantage in terms of simulation time of the Monte Carlo analysis described in Section 5. However, other 6T-cell parameters, like for instance the word-line voltage margin [32] parameter shown in Table 1, need to be computed with the cell being accessed through pass transistors. In this case, simulations were performed considering all the involved circuitry.

The memory cell was characterized computing some parameters of interest. Table 1 summarizes these results.

**Table 1.** Cell-related parameters of interest.

| Cell-Related Parameter | Value |
| --- | --- |
| Supply Voltage | 1.20 V |
| Static Noise Margin (SNM) | 448 mV |
| Read Static Noise Margin (RSNM) | 173 mV |
| Write Static Noise Margin (WSNM) | 450 mV |
| Word-line voltage Margin (WLVM) | 408 mV |
| Single-cell current consumption | 0.104 nA |
| Single-cell write energy | 4.65 fJ |

In this paper, the SRAM memory is intended to be used as PUF. Therefore, in addition to cell related parameters, other PUF related parameters have been obtained following the usual definitions, which can be found in [5,33], and are summarized in Table 2.

**Table 2.** PUF-related parameters of interest. Their values and ideal values are shown.

| PUF-Related Parameter | Value (Ideal Value) |
| --- | --- |
| Uniqueness | 49.8% (50%) |
| Bit aliasing | 49.2% (50%) |
| Reliability | 97.1% (100%) |
| Uniformity | 49.6% (50%) |

Reliability was obtained over a temperature range of −20 °C to +60 °C. Another parameter that can be calculated is bit error rate (BER), which will be further discussed in Section 4.

## 4. SRAM Start-Up Behavior

This section describes and shows the experimental PUF start-up data obtained. As explained in previous sections, cells can be powered down and started up again, as a result, they will end up at '0' or '1' logic values. This behavior depends on slight cell imbalances caused by random parameter variations originated during the manufacturing process.

Preliminary experimental results showed that, as expected, each cell started up at logic '0' or logic '1' showing a spatial random pattern different in the SRAM block of different chips. It was expected because of the random nature of the start-up bias of each cell. This spatial distinctive random pattern will be further analyzed later in this section, and it is the basis for generating a unique fingerprint for each chip.

After these preliminary results, the experimental start-up probability of each cell was measured. To obtain this, the shut-down and power-up process was repeated 1000 times under nominal conditions.

We define the start-up at logic '1' probability ($SUP_1$) of a cell as the ratio of times that it starts up at 1 ($N_1$) with respect to the total number of start-ups, $N$.

$$SUP_1 = \frac{N_1}{N} \quad (1)$$

In a similar way, the start-up at logic '0' probability ($SUP_0$) can be calculated as the ratio of times that it starts up at 0 ($N_0$) with respect to $N$. It follows that $SUP_1 + SUP_0 = 1$.

$$SUP_0 = \frac{N_0}{N} = 1 - SUP_1 \quad (2)$$

In this paper, the $SUP_1$ parameter will be primarily used.

Once the probability of each cell to start up at logic '1' was measured, it was possible to represent in a histogram the relative frequency of cells as a function on $SUP_1$. These results are shown in Figure 3a.

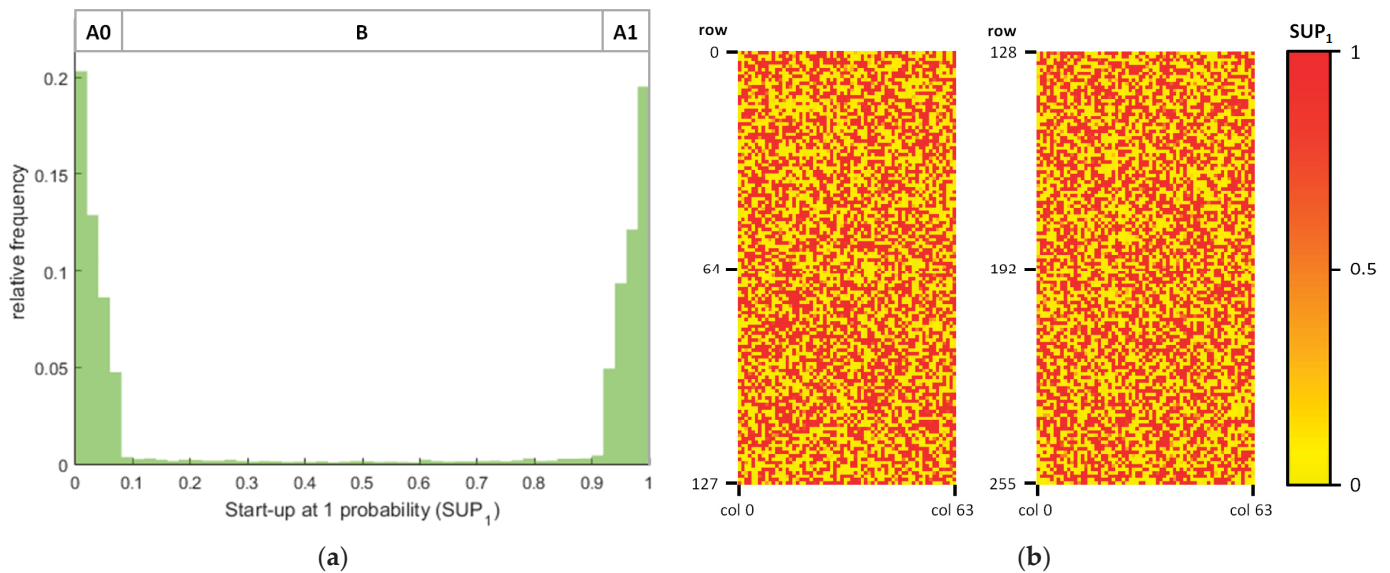

**Figure 3.** (**a**) Histogram of the relative frequency of cell as a function of the $SUP_1$ (start-up at logic '1' probability) obtained from experimental data. Three different regions (A0, B, and A1) can be established. (**b**) Representation of the $SUP_1$ for each different cell in the cell matrix (256 rows × 64 columns) as a function of its location. The matrix is split in two parts for visualization purposes only.

Figure 3a shows that most cells have a preferred logic value to start up. These cells are located at both ends in the horizontal axis of the graph. Note the abrupt change in relative frequency of occurrence around 0.09 and 0.91 $SUP_1$ values. Probabilities very close to 1 mean that the cell is very likely to start up at logic '1', whereas probabilities very close to 0 values mean that the cell is not likely to start up at logic '1', and thus very likely to start up at logic '0'.

By contrast, only a few cells show medium probability and, thus, have an undefined start-up value. These cells are located in the middle (and flat) part of the histogram in Figure 3a.

This allows classifying cells into two different categories: (i) cells with a high probability of starting up at a known value (cells located in both extremes of Figure 3a, in regions A0 and A1). These cells are the most reliable from a PUF application and are a 92% of the total number of cells; and (ii) cells without a clear tendency to start up to a known value (cells located in the middle of Figure 3a, in region B). These cells are unreliable from a PUF point of view and are 8% of the total.

Figure 3b shows the spatial distribution of the $SUP_1$ value of each cell over the entire SRAM floor plan (256 rows × 64 columns). For visualization purposes, the floor plan is split in two parts (from rows 0 to 127, and from rows 128 to 255). Cells with a $SUP_1 = 1$ are represented in pure red, and cells with a $SUP_1 = 0$ are represented in pure yellow. Cells with $SUP_1$ values between 0 and 1 are colored using a mixture of red and yellow, with only those few cells with a $SUP_1 = 0.5$ being orange. Observe that most cells are reddish or yellowish since there are a few cells in region B of Figure 3a.

Additionally, it can be seen that the pattern of $SUP_1$ is spatially random and shows no location dependency. Matrixes from other SRAM parts also show a completely random pattern but the pattern is different from the one shown in Figure 3b. By contrast, the probability distribution of $SUP_1$ is essentially equal from memory to memory. For this reason, even knowing the probability distribution of $SUP_1$, it is not possible to predict the actual cell start-up values in a given sub-set of memory cells, which is the basis for a good PUF. This fact has been previously quantified in Table 2, for instance with the uniqueness parameter, which shows a value very close to the ideal one.

From these results, it turns out that the main issue with this SRAM PUF is not its uniqueness or its bit aliasing, but its reliability: the fact that some cells start up to a value that it is not fully repeatable in time. Bit error rate (BER) is defined as the average ratio of bit errors occurring each time PUF is challenged. We can compute BER of a single cell as the percentage of times that it will not start up to its preferred start value. For instance, a cell with a $SUP_1 = 0.99$ will have a bit error rate of 0.01, and a cell with a $SUP_1 = 0.01$ will have a $SUP_0 = 0.99$, and thus also a bit error rate of 0.01. If we compute the average value over all cells, we obtain 0.045 (4.5%). Obviously, this is an average: most cells will be very reliable and have a much better parameter, and only a few cells will have a parameter close to 50%, which are the undesired cells that can cause problems in the PUF. Consequently, it is important to estimate in the early stages of a design how many of these cells will be present, which is one of the goals of this paper.

Furthermore, it is worth remarking that if error correction codes are used, the definition of a reliable cell can be more or less relaxed depending on the minimum cell reliability needed for the code to meet design specifications.

## 5. Cell Imbalance Factor

Unfortunately, the distribution of start-up values can only be obtained experimentally, that is, once the memory has been manufactured. However, having information about the percentage of reliable cells at the initial design stages is a must. Therefore, it will be very useful to have an alternative parameter that can be obtained from electrical simulation, and whose resulting distribution can be related to the start-up values distribution. This parameter is introduced in this section.

As mentioned in Section 2, each SRAM cell has two stable states: $(V_{DD}, 0)$ and $(0, V_{DD})$. We can represent them in a state space diagram [34] as depicted in Figure 4a, in which each point represents the voltages of the internal nodes $(V_Q, V_{QB})$. The state S0 corresponds to $(V_{DD}, 0)$, and represents the state in which the cell stores a logic '0'; and S1 corresponds to $(0, V_{DD})$, and represents the state in which the cell stores a logic '1'. M is a metastable point. If the cell is perfectly symmetric, this metastable point belongs to the line that goes from $(0, 0)$ to $(V_{DD}, V_{DD})$, and is located at $(V_M, V_M)$, where $V_M$ depends on the relative strength between the pull-up and pull-down devices. This diagonal line is called the separatrix and divides the diagram into two regions, each one related to one of the stable states.

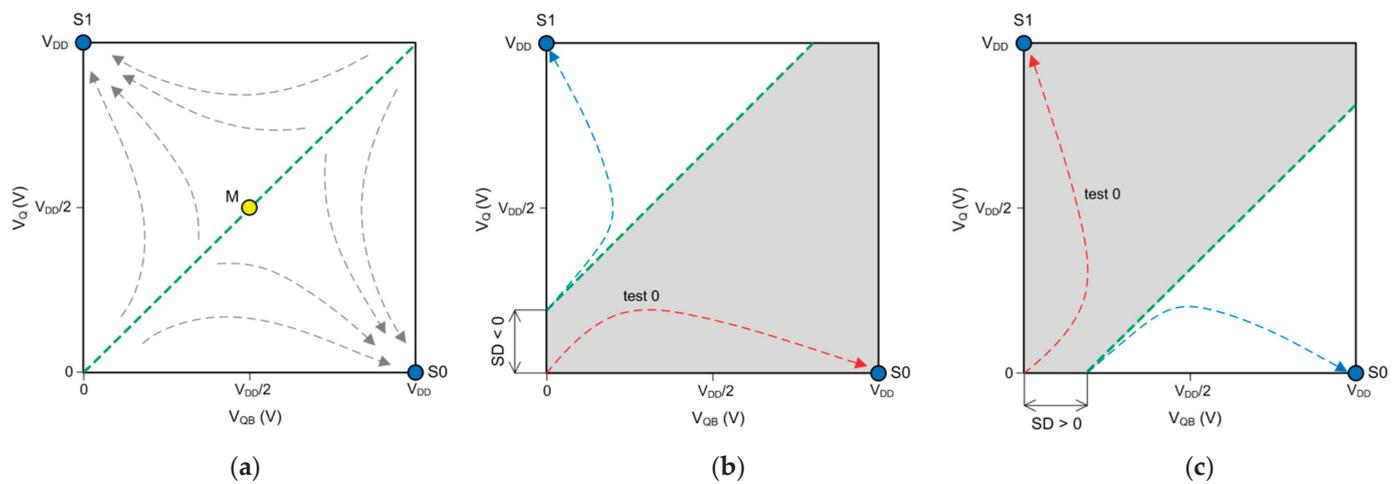

**Figure 4.** State space diagrams for: (**a**) a perfectly symmetric cell; (**b**) a cell biased towards logic '0' state; (**c**) a cell biased towards logic '1' state. Separatrix is represented in green. Each red trajectory (test 0) represents the initial transient analysis used to determine the cell bias, and each blue line is the trajectory with less SD voltage shift that makes the cell start up to the opposite state determined with test 0.

Each of these regions represents the area of attraction of one of the two stable points. This means that if the cell is started at a specific point inside the area, the cell will end up at its corresponding area stable point. This is represented by the grey trajectories depicted in Figure 4a.

Memories are usually started up from the (0, 0) point. In the case of a perfectly balanced cell, this point is on the separatrix, which is also the boundary between the areas of attraction. In this case, the cell will theoretically go to the metastable point and will remain there. Nevertheless, any external or internal perturbation will affect it and drive the cell to one of the two stable states.

However, any real cell will present a mismatch to a certain extent, and thus will be biased towards one of the stable states, as represented in Figure 4b,c. In these cases, the separatrix is not defined by the points (0, 0) and ($V_{DD}$, $V_{DD}$), and thus it divides the phase state into two unequal regions. If the cell is started from (0, 0), it will end up in the stable state of the shadowed area of Figure 4b,c.

The greater the mismatch of a cell, the greater the distance from the point (0, 0) to the separatrix and, consequently, the greater the probability of the cell starting always at the same logic value [35].

So, we can define a cell imbalance factor named SD (Separatrix Distance), which is related to the cell mismatch and can be regarded as a measure of the tendency that a cell has to start up consistently to a given value. Figure 4b,c show this distance labeled "SD". Note that the distance is defined along the axis that the separatrix intersects. A large distance means that the cell will be able to sustain large perturbations without changing its start-up value, and thus the cell will be reliable for PUF applications. On the other hand, a small distance means that a small perturbation will be capable of flipping the preferred start-up state, and thus that the cell will be less reliable from a PUF point of view.

For convenience, we define the SD parameter as positive if the cell is biased towards S1 (logic '1' start-up, like in Figure 4c), and negative if it is biased towards S0 (logic '0' start-up, as shown in Figure 4b).

The computation of the SD parameter was performed using transient Monte Carlo electrical simulations with the Spectre simulator on a commercial 65-nm CMOS technology. The procedure used is as follows:

To determine the state towards the cell is biased to (red trajectories labeled as test 0 in Figure 4b,c), a transient analysis is performed with the cell starting from (0, 0). This simulation indicates which one of the axes it is necessary to explore to compute the SD parameter. For instance, if the cell is biased towards S1 (Figure 4c), we have to search for a starting point ($V_{QB0}$, 0) on the horizontal axis that makes the cell flip its starting-up state to S0.

This has been automated by means of a search algorithm using the Cadence Ocean scripting framework. The algorithm will try different starting points in the axis of interest, and determine the starting point with the minimum voltage that makes the cell start up to the opposite state it was biased to (flip its nominal start-up state) [35]. This is represented by blue trajectories in Figure 4b,c. The procedure is performed over all memory cells.

To sum up, a perfectly symmetric cell will have an SD parameter equal to zero, a cell biased towards logic '1' will have a positive SD, and a cell biased towards logic '0' will have a negative SD parameter.

Since cells are designed to be as symmetric as possible, the SD parameter is expected to have zero average value. In addition, only a small number of cells in a chip are expected to present high imbalance factors, since SRAM are circuits in which design for manufacturing techniques are applied with special care to minimize parameter variations.

The SD parameter was computed for a total of 16,384 6T cell Monte Carlo iterations. This number was chosen because it is the number of cells that are available in the memory described in Section 3.

From the simulation data, the histogram of Figure 5 was computed. It shows the relative frequency of cells as a function of the cell SD parameter.

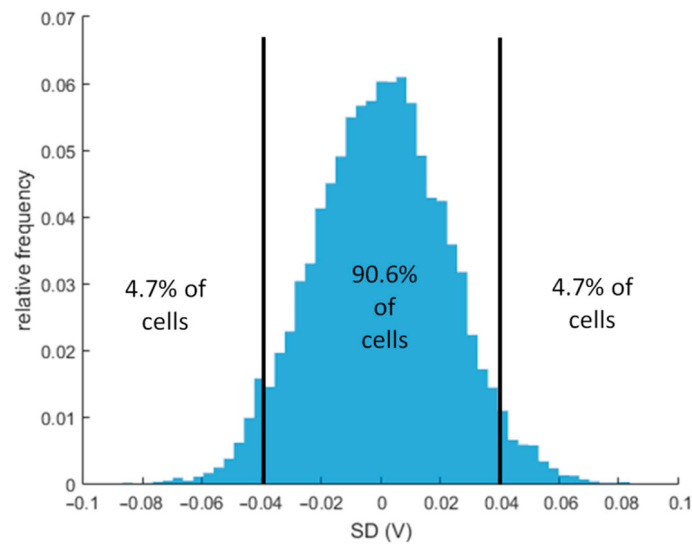

**Figure 5.** Histogram of the relative frequency of cells as a function of the SD parameter obtained from Monte Carlo analysis.

Figure 5 shows that the resulting SD distribution is Gaussian-shaped. As expected, most cells show a relative good matching (SD around zero), and only few cells are highly mismatched (high SD absolute value). For instance, 90.6% of the cells lie between $-0.04$ V and 0.04 V and, consequently, only 9.4% of cells have an SD absolute value factor greater than 0.04 V.

As a consequence of the dynamic behavior of SRAM cells, highly mismatched cells are expected to start up consistently to the same logic value (logic '0' or logic '1') with high probability, while low mismatched cells are expected to start up to a given value with less probability.

The key question is determining the SD absolute value from which a cell can be considered as a highly reliable cell, i.e., which is the threshold value ($SD_{th}$) that makes a cell start up, for instance, more than 95%, 98% or 99% of times to the same value.

Figure 6 is obtained from the data of Figure 5, but considering the absolute value of SD and accumulating the values from left to right. Figure 6 depicts the frequency of cells (vertical axis) with an SD absolute value greater than $SD_{th}$ (horizontal axis). For instance, if obtaining a successful start-up rate greater than 99% required $SD_{th}$ to be 0.05 V, only 3% of the cells would meet the condition. If $SD_{th}$ had to be 0.04, 9.4% of cells would satisfy the condition. This percentage would increase to 21% if the condition was relaxed to $SD_{th}$ = 0.03 V. This question will be addressed in the next section with experimental data.

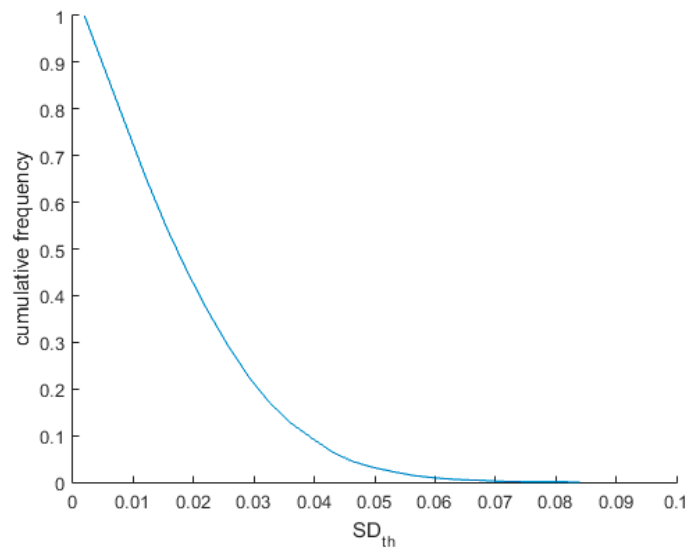

**Figure 6.** Relative number of bit-cells (cumulative frequency on the vertical axis) that present a value of |SD| equal or greater than $SD_{th}$ (horizontal axis).

## 6. Start-Up Value Probability vs. SD Relation

The aim of this section is to find a relation between the simulation values of cell imbalance (*SD* parameter described in Section 5) and the start-up probabilities based on experimental measurements ($SUP_1$ parameter described in Section 4). This will allow determining the threshold value of the *SD* parameter for a cell to start up to a fixed value with a probability higher than the desired target value. This threshold value can be very useful information at the early design stages, since *SD* can be easily obtained by simulation, and because it allows determining the percentage of cells that will be suitable to be used in the target PUF application of the design.

In Section 4, we showed experimentally that most cells (about 92%) started up to a given value with a probability greater than 91%. This means that not only cells with a high *SD* absolute value, but also cells with a moderate *SD* parameter will start up to a given logic state with relatively high probability.

It is important to note that the cell distributions we are dealing with are probability distributions. This means that it is possible that a single cell with very high *SD* absolute value does not always start up to the same value. However, cells belonging to this subset of cells are the ones that are more probable to perform reliably regarding its start-up value.

The correspondence between the probability distribution of the experimental start-up values ($SUP_1$) in Figure 3a and the probability distribution of the *SD* parameter in Figure 5 leads to a series of data-points, which must be fitted with an analytical function. The input of this function is *SD* and it provides the $SUP_1$ as the output. This function needs to be chosen and meet the following requirements:

1. *SD* = 0 must lead to $SUP_1$ = 0.5 (start-up at logic '1' probability equal to 0.5).
2. $SD \gg 0$ V must lead to $SUP_1$ = 1

3. $SD \ll 0$ V must lead to $SUP_1 = 0$, and thus $SUP_0 = 1$

These requirements are in principle satisfied by the logistic function, a common S-shape curve (sigmoid curve) that has been extensively used to model research problems in many fields [36]. Its expression in terms of $SUP_1$ and $SD$ is given by:

$$SUP_1 = \frac{a}{1 + e^{-k\,SD}} \tag{3}$$

where $a$ is the curve maximum value when $SD \gg 0$ V, and $k$ is the curve steepness parameter. So, in our case, due to the second requirement, $a = 1$, and $k$ must be determined by fitting with the experimental $SUP_1$ data. A logistic curve with $a = 1$ is shown in Figure 7a.

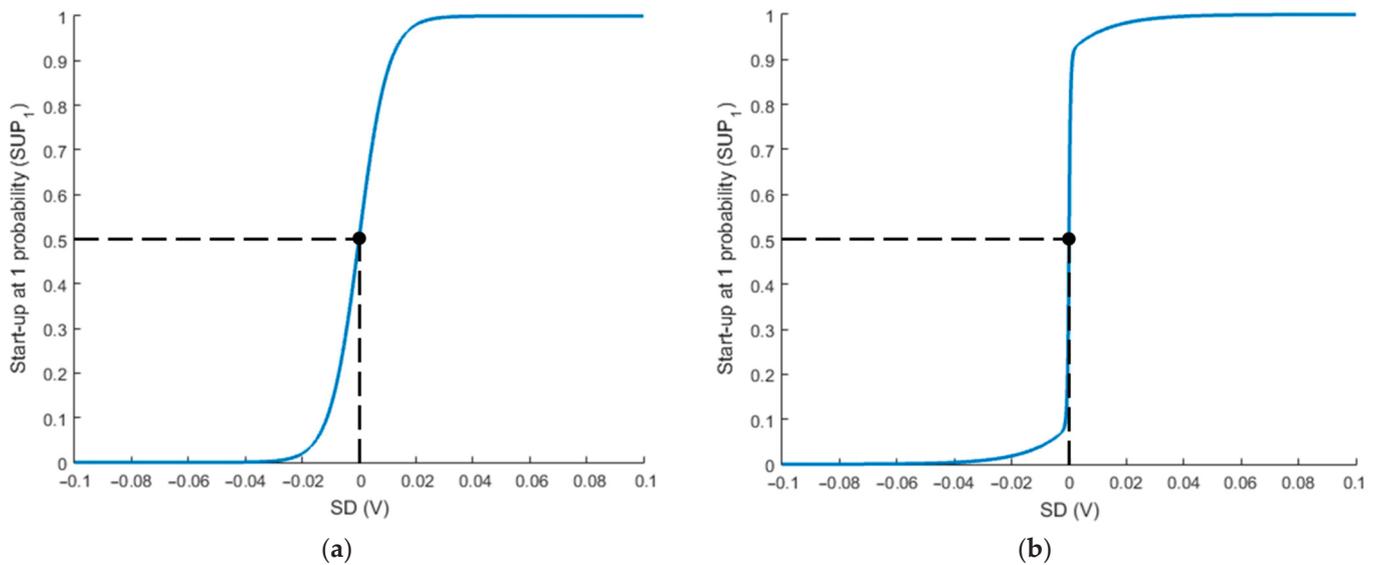

**Figure 7.** (**a**) Plot of the sigmoid function defined in Equation (3) after a fitting process (a = 1; k = 195.4 V$^{-1}$); (**b**) Sigmoid function that meets the requirements of the proposed model. The plot corresponds to the function defined in Equation (5) after a fitting process (m = 0.158; k$_1$ = 101.2 V$^{-1}$; k$_2$ = 2348 V$^{-1}$).

The fitting process tries to fit the simulated $SUP_1$ distribution, obtained from Equation (3) when using the $SD$ values resulting from the Monte Carlo analysis, with the experimental $SUP_1$ distribution. When the fitting was performed, the following result was obtained:

$$k = 195.4\ V^{-1} \tag{4}$$

The sigmoid curve in Figure 7a plots Equation (3) with $a = 1$ and this fitted parameter value.

Figure 8a shows the relative frequency of cells as a function of $SUP_1$ for both the experimental data and the simulated data once transformed into start-up probabilities using Equation (3) and fitted. As can be seen, the matching between the two sets of data is not good enough. This is because the experimental data (Figure 3a) show an abrupt transition around 0.09 and 0.91 $SUP_1$ values, which a function with only one fitting parameter is not capable of correctly modeling.

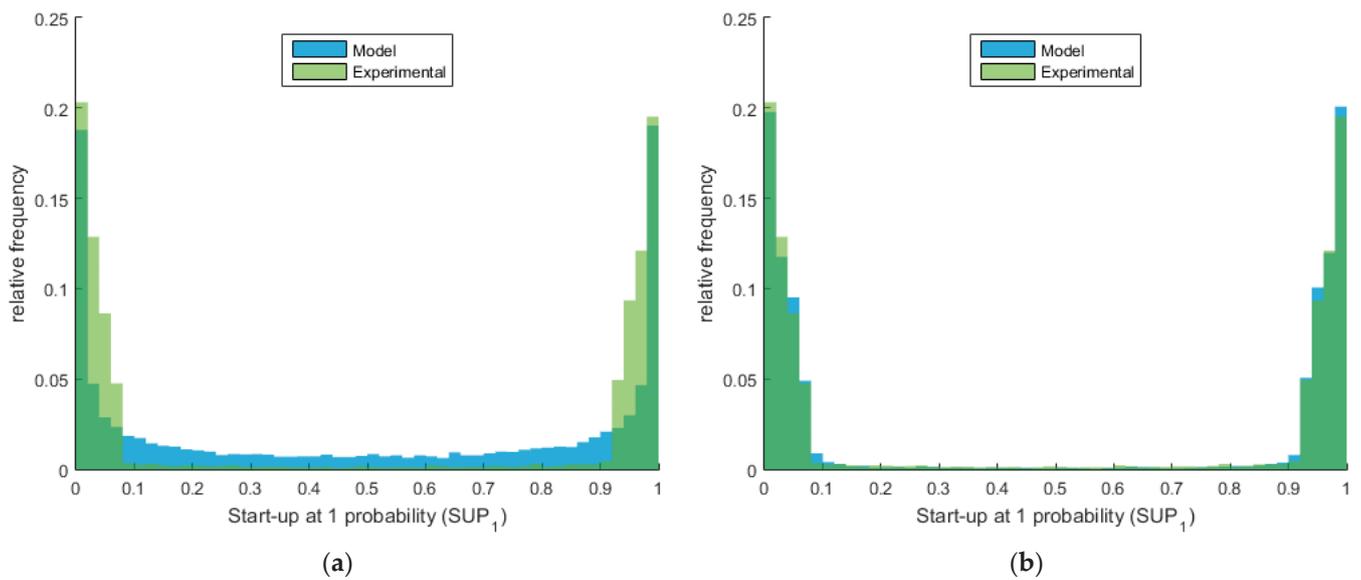

**Figure 8.** Histogram of the relative frequency of cells as a function of the $SUP_1$ (start up at logic '1' probability) obtained from both experimental data and from modeled simulation results using: (**a**) Equation (3) with the fitted parameter of Equation (4) ($k$ = 195.4 V$^{-1}$), and $a$ = 1; (**b**) Equation (6) with the fitted parameters of Equation (7) ($m$ = 0.158; $k_1$ = 101.2 V$^{-1}$; $k_2$ = 2348 V$^{-1}$).

To improve the modeling of the transition between the different regions, a function with 2 logistic components is considered:

$$SUP_1 = \frac{m_1}{1 + e^{-k_1\, SD}} + \frac{m_2}{1 + e^{-k_2\, SD}} \quad (5)$$

Each component has its own exponential coefficient ($k_1$ and $k_2$) and its own weight ($m_1$ and $m_2$). To meet the requirements, it is necessary that $m_1 + m_2 = 1$. So, we impose that $m_1 = m$, and $m_2 = 1 - m$, and Equation (5) can be written as:

$$SUP_1 = \frac{m}{1 + e^{-k_1\, SD}} + \frac{1-m}{1 + e^{-k_2\, SD}} \quad (6)$$

Observe that the new weights add to 1. As a consequence, it can be easily proved that when $SD \gg 0$, $SUP_1$ tends to 1; when $SD \ll 0$, $SUP_1$ tends to 0; and $SUP_1(0) = 0.5$, so it meets the requirements.

A representation of this function can be seen in Figure 7b, where it can be observed that the curve meets all of the requirements and that the function shows odd symmetry with respect to the (0, 0.5) point. This is a consequence of the symmetry of both $SUP_1$ (Figure 3a) and SD (Figure 5) distributions.

Parameters $m$, $k_1$ and $k_2$ must be determined by fitting the simulated $SUP_1$ distribution, obtained from the Equation (6) when using the *SD* values resulting from the Monte Carlo analysis, to the experimental $SUP_1$ distribution.

After completing the fitting, the following result has been obtained:

$$m = 0.158 \quad k_1 = 101.2\ V^{-1} \quad k_2 = 2348\ V^{-1} \quad (7)$$

The curve in Figure 7b plots Equation (6) with the fitted parameter values. Figure 8b shows the relative frequency of cells as a function of $SUP_1$ for both the experimental data and the simulated data converted into start-up probabilities using Equation (6) and properly fitted. As can be seen, unlike the fitting with Equation (3), now with Equation (6) the agreement between the two sets of data is good.

Notice that the two exponential constants (Equation (7)) show very different values, which is a consequence of modeling the two different regions (A and B) labeled in Figure 3a. In Figure 7b, this can also be seen as two different regions: one near *SD* = 0, which has a very steep slope, and the regions far away *SD* = 0, where the slope decreases and eventually becomes zero. In fact, if the slope of the curve of Figure 7b near *SD* = 0 is calculated, it gives approximately 500 V$^{-1}$, which roughly means that increasing *SD* by 0.2 mV rises the start-up at '1' probability by 0.1. This result is a numeric quantification of the fact that small differences in cell unbalance lead to great differences in start-up probability. For this reason, there are few cells that start up at logic '1' with near 0.5 probability, because small differences of only a fraction of mV in the *SD* parameter are translated into great differences in start-up probability. This means that even slightly asymmetric cells are quite biased towards one specific start-up value, and thus can be good candidates for PUF.

From the curve depicted in Figure 7b, it can be calculated that absolute values of SD greater than 0.03 V lead to start-up at '1' probabilities higher than 0.99 (and by symmetry the same happens for start at '0'). Finally, the question set out in Section 5 can be answered: 21% of cells present an SD$_{th}$ = 0.03 V and these cells have a probability greater than 99% of maintaining the same start-up value.

Table 3 summarizes the results of SD$_{th}$ and the percentage of cells for three different probabilities of the cell to repeat start-up value.

**Table 3.** SD$_{th}$ parameter value and percentage of cells for different repeating start-up value probabilities.

| Repeating Start-Up Value Probability | SD$_{th}$ (V) | Percentage of Cells |
| --- | --- | --- |
| 0.99 | 0.030 | 21% |
| 0.98 | 0.019 | 42% |
| 0.95 | 0.008 | 76% |

We obtained the PUF reliability [5] parameter to take into account the effect of start-up time and temperature. The considered temperature range was −20 °C to +60 °C and the start-up time was varied more than one order of magnitude according to [37,38]. Results were 97.1% and 98.3% respectively, suggesting that the percentages of cells listed in Table 3, which as stated by [35] correspond to the least influenced by external disturbances, will be largely unaffected by these factors. These results are in line with those of [35,39].

## 7. Conclusions

We proposed a method to predict the statistical start-up probability distribution of a given SRAM block not specifically designed for PUF applications. The method firstly computes, using Monte Carlo analysis, the distribution of a cell imbalance factor, SD, based on the distance from the origin point of the 6T-cell space state diagram to the separatrix. Then, SD is transformed to a start-up probability distribution applying a transfer function that has been calibrated with experimental results.

The experimental results show that, for the analyzed SRAM implemented with a 65-nm CMOS technology, around 21% of cells will start at the same logic value with a probability higher than 99%. Based on the hypothesis that those cells that present higher SUP values are the same ones that present higher SD factor values, we deduce that a bit-cell with SUP$_1 \geq$ 99% corresponds to a cell with SD $\geq$ 0.03 V.

Generalizing, the described method allows the designer to predict, using Monte Carlo results from electrical simulation, the proportion of cells that will start up to the same logic value with a given probability or, in other words, the number of cells of a specific SRAM block that will be reliable for a particular PUF application.

The prediction can be obtained at the early design stages. Depending on the prediction results, the design team could decide whether the SRAM meets the specifications. If they are not met, they could choose either using a larger memory in order to have more cells, or apply error correction techniques and evaluate its requirements to achieve the overall desired PUF performance.

**Author Contributions:** Conceptualization, G.T. and S.A.B.; methodology, G.T. and S.A.B.; validation, G.T., A.A. and B.A.; investigation, G.T. and A.A.; resources, G.T. and S.A.B.; data curation, G.T. and A.A.; writing—original draft preparation, G.T.; writing—review and editing, G.T., B.A. and S.A.B.; supervision, G.T., B.A. and S.A.B.; funding acquisition, G.T. and S.A.B. All authors have read and agreed to the published version of the manuscript.

**Funding:** This research received no external funding.

**Institutional Review Board Statement:** Not applicable.

**Informed Consent Statement:** Not applicable.

**Conflicts of Interest:** The authors declare no conflict of interest.